\documentclass[11pt,a4paper]{article}
\usepackage[T1]{fontenc}
\usepackage[utf8]{inputenc}
\usepackage{lmodern}
\usepackage{amsmath,amssymb}
\usepackage[margin=22mm]{geometry}
\usepackage{graphicx}
\usepackage{booktabs}
\usepackage{array}
\usepackage{tabularx}
\usepackage{longtable}
\usepackage{ragged2e}
\usepackage{placeins}
\usepackage{microtype}
\usepackage[hidelinks]{hyperref}
\newcolumntype{Y}{>{\RaggedRight\arraybackslash}X}
\newcolumntype{P}[1]{>{\RaggedRight\hspace{0pt}\arraybackslash}p{#1}}
\title{Generating is not discovering: a pre-registered physics judge for AI-proposed superconductors, calibrated on six known superconductors and one negative control}
\author{Reinaldo In\'acio\\[2pt] \small \href{https://orcid.org/0009-0000-6594-6014}{ORCID 0009-0000-6594-6014}\\ \small BITA (Inova Simples), S\~ao Paulo, Brazil \quad \texttt{reinaldo.inacio@bitatech.com.br}}
\date{8 September 2026}
\begin{document}
\maketitle

\begin{abstract}
Generative models now propose millions of ``stable'' crystal structures, and their own referees have shown that stability is not discovery; no public benchmark asks whether a candidate has the physics to do a job. For superconductors the job is twofold: to pair, with a definite gap symmetry, and to become wire. We present a physics judge that answers both from a structure alone (DFT+U, a gated Wannier model, the full rank-4 RPA susceptibility on a $24^{3}$ mesh, the linearized gap equation, an irreducible-representation classifier, a manufacturability funnel) and measure what it says about what generators propose. With answer keys pre-registered before computing, the judge was calibrated on six known superconductors (Nb, Nb$_{3}$Sn, MgB$_{2}$, BaFe$_{2}$As$_{2}$, La$_{2-x}$Sr$_{x}$CuO$_{4}$, YBa$_{2}$Cu$_{3}$O$_{7}$), recovering the expected class in all six after two dated corrections. On the ruthenium analogue of the iron pnictide, with two runs and their criterion pre-registered the day before, it returns a null verdict at the same physical interaction where iron gives s$\pm$ ($\lambda_{1}$ $\le$ 0.0007 versus 0.096; critical U 4.17 versus 0.97 eV); the pre-registered equal-$\alpha$ run fails the letter of the criterion, and we declare the resolution in favour of absolute U as post-hoc. The funnel reproduces the industrial map with thresholds fixed before running. Three audits of generation follow: a descriptor sweep over 150 Materials Project metals re-finds the canon; 1,248 MatterGen structures yield 0 candidates that are new, stable and carry a pairing motif; mechanism filters over 47,893 compounds, validated by a blind hold-out, re-derive the community's analogies: geometry plus d-count is necessary but not sufficient. Generation re-finds what is known; judgment is the bottleneck. We propose a public benchmark over the eleven public lists of AI-proposed superconductor candidates, none of which classifies gap symmetry. Total cloud cost: under US\$ 200.
\end{abstract}

\section{Introduction}
Three facts frame this work. First, generation has become cheap: GNoME reported 2.2 million structures and released 380,000 to the Materials Project [1]; MatterGen generates structures conditioned on chemistry and properties [2]; companies built around generation raised between US\$ 30 million and US\$ 450 million per round in 2024--2026 [3--5]. Second, the field's referees have shown that generated ``novelty'' is mostly not: Leeman, Schoop, Palgrave and co-workers found that about two thirds of the 43 compounds the autonomous A-Lab claimed as new were likely known, compositionally disordered versions of the predicted ordered phases, and that ``no new materials have been discovered'' in that campaign [6]; Cheetham and Seshadri found ``scant evidence'' that GNoME's compounds are simultaneously novel, credible and useful [7]; Nature published an Author Correction to the A-Lab paper on 19 January 2026 [8]. Third, the public benchmark that ranks generators, Matbench Discovery, measures thermodynamic stability [9]. There is no public benchmark that asks the question a superconductor programme needs answered: does this candidate pair, with which symmetry, and can it be drawn into wire?

That question is where the discovery of a manufacturable superconductor is actually stuck. YBa$_{2}$Cu$_{3}$O$_{7}$ has existed since 1987 and its tape still costs about US\$ 300 per kA$\cdot$m at the operating conditions of compact-fusion magnets, against a target near US\$ 10 [10], because a d-wave gap makes the critical current collapse at grain boundaries and forces epitaxial, biaxially textured conductors. The symmetry of the gap decides the route to wire before any furnace is lit. A judge that classifies it, and that says \emph{no} when the physics is absent, is the layer that sits between generation and synthesis and for which no public benchmark exists, because it sits between communities: machine-learning groups do not run many-body calculations on a thousand candidates, and many-body physicists do not run a thousand candidates.

This paper does three things. (i) It describes a judge that runs end-to-end from a structure, with coded gates at every stage and the complete configuration embedded in every result. (ii) It reports the judge's calibration on six known superconductors with answer keys pre-registered before the calculations, one negative control with success and failure criteria written the day before it ran, and one campaign the author hoped would succeed and closed as a failure with its cause. (iii) It measures, with three independent routes, what generation alone delivers in the relevant chemical families, and proposes the public benchmark that the field lacks.

\section{The judge}
\begin{figure}[t]\centering\includegraphics[width=\textwidth]{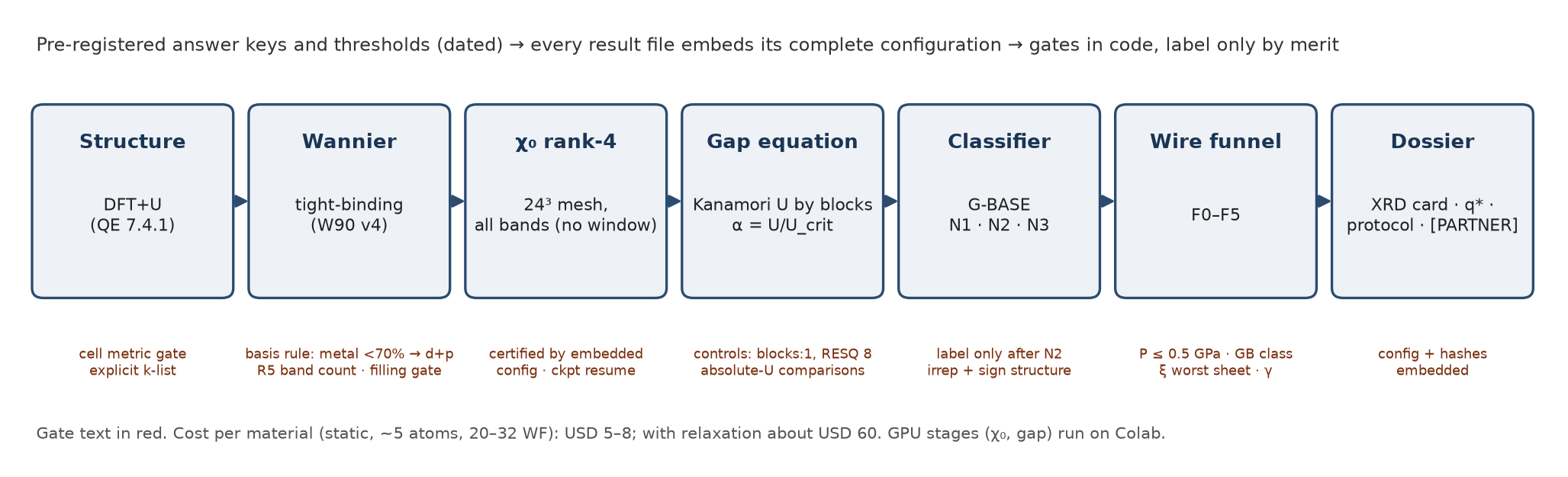}\caption{The judge: seven stages and their gates. Answer keys and thresholds are pre-registered and dated; every result file embeds its complete configuration; a label is issued only by merit.}\label{fig:pipeline}\end{figure}
\subsection{Chain}
The pipeline (\texttt{bita\_run.py}, manifest-driven, phases \texttt{estrutura $\to$ scf $\to$ nscf $\to$ wannier $\to$ r5 $\to$ chi0 $\to$ trilha2 $\to$ dossie}) takes a crystal structure and produces a dossier. The stages and their gates:

\begin{enumerate}
\item \textbf{DFT+U} (Quantum ESPRESSO 7.4.1, PAW, GGA) on the experimental or relaxed cell; a structure gate checks the cell metric against the declared target; the non-self-consistent step uses an explicit uniform k-list (never automatic sampling, which destroys the Wannier disentanglement).
\item \textbf{Wannier tight-binding model} (Wannier90 v4). The basis is decided by a transferability check on the projected character of the active metal around $E_F$, which selects between a metal-only and a metal-plus-ligand (d+p) basis before any model is built. Two gates: the band-tracking validator R5 (exact count of bands in the active window, mean absolute error in meV, ``complete manifold'') and a filling gate that requires the tight-binding model to reproduce the DFT occupation of every band crossing $E_F$, not only its dispersion. The filling gate exists because a model that under-filled the Zhang--Rice band of La$_{2-x}$Sr$_{x}$CuO$_{4}$ produced a spurious irreducible representation.
\item \textbf{Bare susceptibility} $\chi_{0}$(q) as a full rank-4 tensor in the Wannier basis, computed on a $24^{3}$ k-mesh ($12^{3}$ was shown twice to be an artefact regime), over \textbf{all} bands of the model with no energy window (``complete $\chi_{0}$'', certified by the \texttt{janela: null} field of the embedded configuration), then projected onto the correlated subspace. Computed in JAX on GPU.
\item \textbf{Linearized gap equation} in the RPA with a Kanamori interaction defined by blocks (the five d orbitals of each metal site), J/U = 0.15. The interaction strength is expressed as $\alpha$ = U/$U_{\mathrm{crit}}$, where $U_{\mathrm{crit}}$ = 1/$\lambda_{\max}$($\chi_{0}\cdot$U$_{1}$) is the RPA instability of the material itself; results are reported at $\alpha$ = 0.90, 0.95, 0.97 and, for comparisons between materials, at the same absolute U (\S{}4.1). Two mandatory controls on the same $\chi_{0}$ checkpoint: a single-block interaction and a coarser q-mesh. All susceptibilities in this work are computed at T = 130 K: a single temperature above the highest $T_c$ in the set (92 K), so that normal-state susceptibilities are compared on the same footing and the thermal broadening is resolved by the $24^{3}$ mesh; every comparison between materials (\S{}3, \S{}4.1) is at this T. Because U is defined through $\alpha$ = U/$U_{\mathrm{crit}}$ of each material, the absolute value of U enters only where two materials are compared at the same U (\S{}4.1); a reader used to cRPA values of 2--4 eV in a d-only basis should read the 0.92 eV of \S{}4.1 as the Kanamori U at which BaFe$_{2}$As$_{2}$ sits at $\alpha$ = 0.95 of its own RPA instability in this basis, not as an estimate of the screened U.
\item \textbf{Symmetry classifier (G-BASE)} with three levels: N1 (invariants of $\chi_{0}$: is there a selected q*, or dust?), N2 (is the gap solution stable across $\alpha$ and controls?), N3 (irreducible representation of the leading eigenvector under the point group of the cell, and the sign structure across Fermi sheets: s++, s$\pm$, d, \ldots{}). A label is issued only when N2 passes (``gate before label'').
\item \textbf{Wire funnel} F0--F5: F0 stability at fabricable pressure ($\le$ 0.5 GPa, or epitaxial strain $\le$ 3 \%); F1 grain-boundary class from N3 (A: nodeless uniform s, round wire; B: s$\pm$ inter-sheet, round wire viable, powder-in-tube precedent; C: nodes of symmetry, coated conductor only); F2 coherence length $\xi_{0}$ = $\hbar\langle$$v_F$$\rangle$/($\pi\Delta_{0}$) on the \emph{worst} Fermi sheet carrying $\ge$ 5 \% of the condensate ($\ge$ 1.5 nm pass, $\ge$ 1.0 warning); F3 anisotropy $\gamma$ from Fermi-surface-weighted velocities ($\le$ 7 pass, $\le$ 20 warning); F4 carriers and F5 Mott distance U/W as flags. Thresholds were pre-registered on 1 September 2026 with one dated amendment (the 5 \% condensate rule, introduced because a shallow pocket with 0.1 \% of the condensate failed YBa$_{2}$Cu$_{3}$O$_{7}$, a real coated-conductor material).
\item \textbf{Dossier}: crystallographic card with the theoretical Cu K$\alpha$ diffraction pattern, what spectroscopy must see (density of states, Fermi sheets, susceptibility peak q*), the gap class with a confirmation and refutation protocol, the funnel verdict, and the fields the pipeline does not compute (synthesis route, kinetics, disorder) left explicitly to the experimental partner.
\end{enumerate}
\subsection{What the judge is not}
It is a weak-coupling, spin-fluctuation judge. It does not compute phonons beyond a density-of-states gate that stops phononic materials early (\S{}3); it does not relax structures unless asked; it does not compute defect thermodynamics or synthesizability. It is designed to be wrong in a detectable way: every result file embeds $E_F$, U, J/U, the orbital indices, both meshes, the window, the hash of the tight-binding model and the code version, so any verdict can be re-run.

\subsection{Cost}
Following the written setup manual, a new material of \textasciitilde{}5 atoms and 20--32 Wannier functions costs US\$ 5--8 of cloud compute end-to-end (static cell), \textasciitilde{}US\$ 60 with structural relaxation. The six-material calibration suite cost about US\$ 4; the negative control of \S{}4.1, including three Wannier rounds that became rules, about US\$ 30; the strain campaign of \S{}4.2, US\$ 53.

\subsection{Pre-registration protocol}
Before each computation a dated file states the expected class, the criterion that would count as failure, and the action in each case. The calibration answer keys (28 August 2026), the funnel thresholds (1 September), the negative-control criteria and the rule for the interaction (5 September), and the mechanism-filter tolerances with their blind hold-out (5 September, three dated lines) are all in the repository. Where the result contradicted the expectation, the contradiction is reported (\S{}3, flags; \S{}4.2).

\section{Calibration on six known superconductors}
\begin{figure}[t]\centering\includegraphics[width=\textwidth]{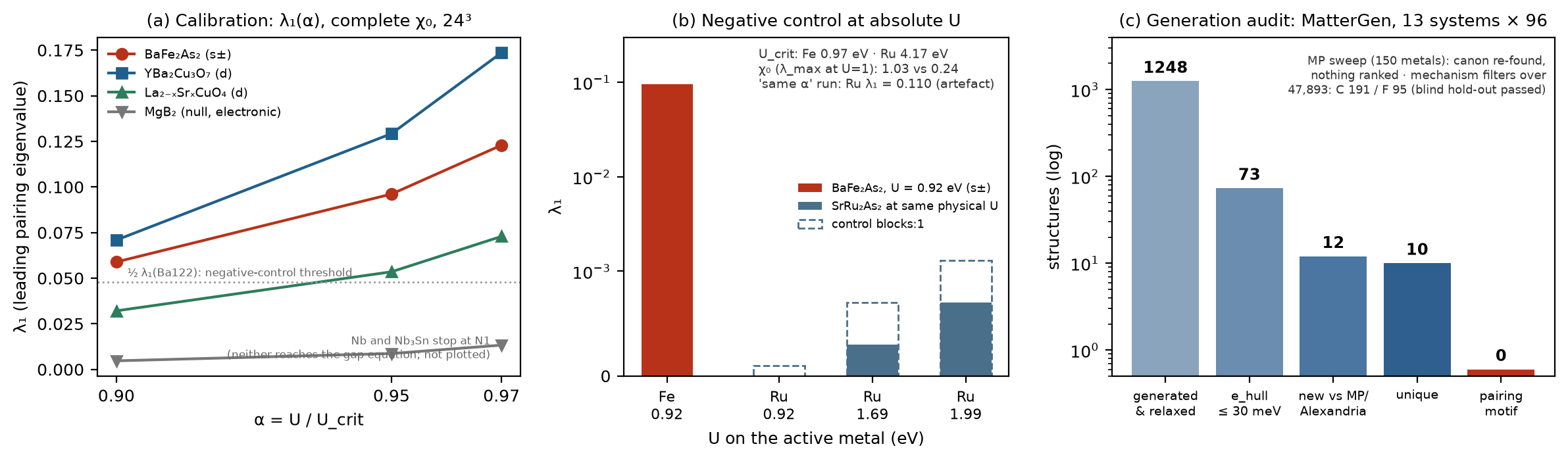}\caption{(a) Calibration: leading pairing eigenvalue $\lambda_{1}$ versus $\alpha$ = U/$U_{\mathrm{crit}}$ for the four materials that reach the gap equation, all with complete $\chi_{0}$ on a $24^{3}$ mesh at T = 130 K (Table S1); Nb and Nb$_{3}$Sn stop at the N1 gate, so neither is plotted. (b) Negative control at absolute U: BaFe$_{2}$As$_{2}$ versus SrRu$_{2}$As$_{2}$ with the single-block control; the equal-$\alpha$ run ($\lambda_{1}$ = 0.110) is the artefact discussed in \S{}4.1. (c) Generation audit: the pre-registered MatterGen funnel, 1,248 $\to$ 0.}\label{fig:validation}\end{figure}

\begin{figure}[t]\centering\includegraphics[width=\textwidth]{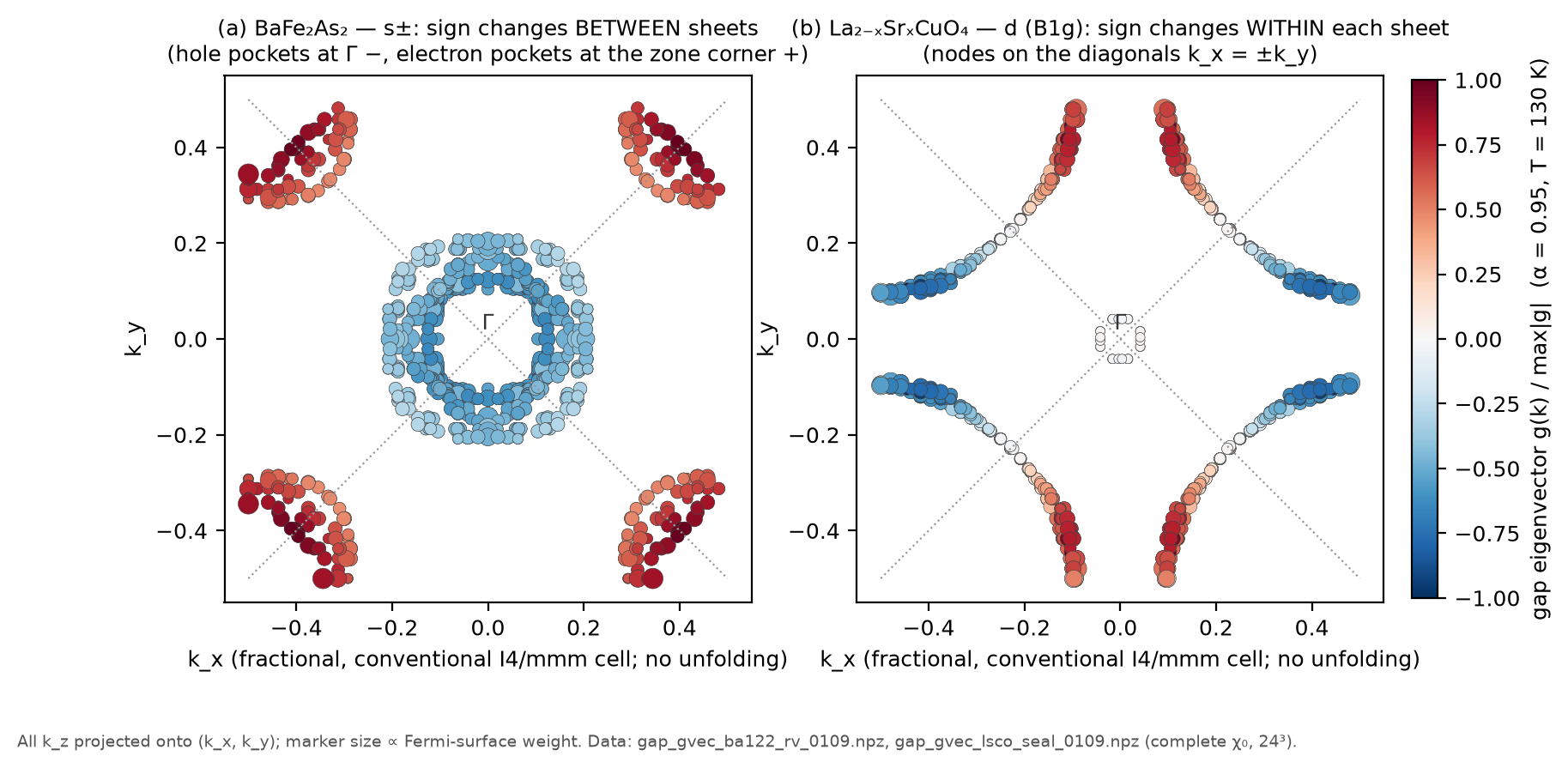}\caption{Leading gap eigenvector on the Fermi surface, $\alpha$ = 0.95, T = 130 K, plotted in the Brillouin zone of the conventional I4/mmm cell used by each tight-binding model, with all $k_z$ projected onto ($k_x$, $k_y$) and no unfolding applied. (a) BaFe$_{2}$As$_{2}$, 2-formula-unit cell (4 Fe; a = 3.9625 \AA{}, c = 13.017 \AA{}): the sign changes between the hole pockets at $\Gamma$ and the electron pockets at the zone corner (s$\pm$). (b) La$_{2-x}$Sr$_{x}$CuO$_{4}$, 2-Cu cell (a = 3.78 \AA{}, c = 13.18 \AA{}; second Cu at $\tau$ = ($\tfrac{1}{2}$, $\tfrac{1}{2}$, $\tfrac{1}{2}$)), so the two folded sheets are the same CuO$_{2}$ band: the sign changes within each sheet with nodes on the diagonals (d, B1g). Marker size is proportional to Fermi-surface weight.}\label{fig:gap}\end{figure}

{\small\setlength{\tabcolsep}{3pt}
\begin{longtable}{P{23mm}P{15mm}P{32mm}P{15mm}P{34mm}P{33mm}}
\caption{Calibration on six known superconductors: verdict, answer key, matching detail and declared flag per material.}\label{tab:calibration}\\
\toprule
\textbf{material} & \textbf{regime} & \textbf{judge's verdict} & \textbf{answer key} & \textbf{detail that matches} & \textbf{flag declared} \\
\midrule
\endfirsthead
\multicolumn{6}{l}{\small Table~\ref{tab:calibration} (cont.)}\\
\toprule
\textbf{material} & \textbf{regime} & \textbf{judge's verdict} & \textbf{answer key} & \textbf{detail that matches} & \textbf{flag declared} \\
\midrule
\endhead
\midrule\multicolumn{6}{r}{\small continued on next page}\\
\endfoot
\bottomrule
\endlastfoot
Nb & conventional & N1 null: q* at a generic point of the zone, the ten strongest q within 1 \% of each other, no selected wave vector; the same null on two bases (Nb-d, 5 WF, 28 Aug 2026; Nb s+d, 6 WF, 7 Sep 2026) & phononic & band tracking exact on both models (MAE 0.014 meV; 13 $\mu$eV inside the frozen window of the s+d model); filling gate passed on the same Hamiltonian as $\chi_{0}$ (s+d: $\Delta$ = 0.029, exact on the DFT mesh; d-only: $\Delta$ = 0.031, exact on the DFT mesh) & the d-only model carries the occupied s band inside its frozen window (two Wannier functions of 7--9 \AA$^{2}$); the s+d model removes this and returns the same verdict. An earlier evaluation of the gate that discarded the imaginary part of H(R) had failed the d-only model ($\Delta$ = 0.208); it was inconsistent with the Hamiltonian the $\chi_{0}$ uses and is superseded (Table S1) \\
Nb$_{3}$Sn & conventional & N1 null: q* migrates with temperature (argmax noise), no selected q & phononic & 30 Nb-d Wannier functions; band tracking exact (MAE 0.011 meV) and filling gate passed ($\Delta$ = 0.020); the only caveat is one band grazing the edge of the frozen window, which the band-count check flags at the boundary & --- \\
MgB$_{2}$ & conventional ($\sigma$ bands) & \textbf{outside the scope of a spin-fluctuation judge}: no electronic pairing channel, $\lambda_{1}$ = 0.009 with complete $\chi_{0}$ (6 Sep 2026); this is consistent with, not proof of, the phononic mechanism & phononic & $\sigma$ character at $E_F$ 44 \% vs 42 \% in the literature; with the corrected window $\lambda_{1}$ fell 4$\times$, the null got stronger & --- \\
BaFe$_{2}$As$_{2}$ & iron pnictide & \textbf{s$\pm$}, A1g weight 0.999, $\lambda_{1}$ = 0.096 at $\alpha$ = 0.95; hole sheets $-$, electron sheets + & s$\pm$ & q* = ($\tfrac{1}{2}$, $\tfrac{1}{2}$, 0), the wave vector of the real stripe spin-density wave; 10 of 10 pre-registered criteria met & --- (an evaluation of the filling gate that discarded the imaginary part of H(R) had placed this model in the warning zone, $\Delta$ = 0.065; on the Hamiltonian the $\chi_{0}$ uses it passes cleanly, $\Delta$ = 0.021; Table S1) \\
La$_{2-x}$Sr$_{x}$CuO$_{4}$ & cuprate & \textbf{d (B1g)}, weight 1.000; nodes on the diagonals of each sheet & d & t = 0.40--0.44 eV, t$'$/t = $-$0.10 as in the literature; q* = ($\tfrac{1}{2}$, $\tfrac{1}{2}$, \textasciitilde{}0), the ($\pi$, $\pi$) antiferromagnetic vector & the first model (2 WF) under-filled the Zhang--Rice band and returned a spurious irrep; the filling gate and a re-Wannierization under that gate fixed it (dated 30 Aug--1 Sep) \\
YBa$_{2}$Cu$_{3}$O$_{7}$ & cuprate, orthorhombic & \textbf{d}: Ag of D2h with $d_{x^2-y^2}$ planar weight 0.944, s-extended 0.03 & d & prediction committed to the repository before the result; the 3 \% s admixture is what a/b junction experiments see & q* of the full-cell $\chi_{0}$ at ($-\tfrac{1}{4}$, $\tfrac{1}{4}$, $\tfrac{1}{2}$) instead of the expected ($\tfrac{1}{2}$, $\tfrac{1}{2}$, $\tfrac{1}{2}$); a pre-registered plane-projected test (6 Sep) did not close it: ``class correct, mechanism under investigation'' \\
\end{longtable}}
Six of six recover the expected class with one instrument and no per-material tuning, after two dated corrections (the LSCO re-Wannierization and the MgB$_{2}$ window) and one corrected gate evaluation (below); two flags are declared (Nb basis, YBa$_{2}$Cu$_{3}$O$_{7}$ q*). This is calibration accuracy, not blind-test accuracy (\S{}8). All four $\chi_{0}$ results that enter the symmetry chain (BaFe$_{2}$As$_{2}$, La$_{2-x}$Sr$_{x}$CuO$_{4}$, YBa$_{2}$Cu$_{3}$O$_{7}$, MgB$_{2}$) are certified as complete by the embedded configuration (\texttt{janela: null}, $24^{3}$, T = 130 K; Table S1 lists file, mesh and model hash). All six tight-binding models (and the ruthenide of \S{}4.1) pass the Wannier-versus-DFT filling gate when the gate is evaluated on the same Hamiltonian that the $\chi_{0}$ uses, the complex H(R) written by Wannier90 (Table S1; largest deviation 0.031, on the d-only Nb model, with zero deviation on the DFT mesh itself). Table S1 also lists a superseded evaluation, dated 1--7 September 2026, in which the gate discarded the imaginary part of H(R): it had placed BaFe$_{2}$As$_{2}$ and SrRu$_{2}$As$_{2}$ in the warning zone and had failed the Nb model; the discrepancy was found on 7 September while re-Wannierizing Nb and is reported as a corrected gate evaluation, not as a change of physics (the susceptibility and gap results never used the truncated Hamiltonian). \textbf{Figure~\ref{fig:pipeline}} shows the chain and its gates; \textbf{Figure~\ref{fig:validation}a} shows $\lambda_{1}$($\alpha$) for the four materials that reach the gap equation (Nb and Nb$_{3}$Sn stop at N1; neither reaches the gap equation and neither is plotted); \textbf{Figure~\ref{fig:gap}} shows the leading gap eigenvector on the Fermi surface of BaFe$_{2}$As$_{2}$ (sign change \emph{between} the hole pockets at $\Gamma$ and the electron pockets at the zone corner: s$\pm$) and of La$_{2-x}$Sr$_{x}$CuO$_{4}$ (sign change \emph{within} each sheet with nodes on the diagonals: d, B1g), all $k_z$ projected onto ($k_x$, $k_y$).

\section{The judge says no}
\subsection{\texorpdfstring{Negative control: SrRu$_{2}$As$_{2}$, same geometry as the iron pnictide, no relevant superconductivity}{Negative control: SrRu2As2, same geometry as the iron pnictide, no relevant superconductivity}}
The first technical question any reader asks is whether a spin-fluctuation judge approves every square lattice. We chose SrRu$_{2}$As$_{2}$ (ThCr$_{2}$Si$_{2}$ type, a square Ru net in RuAs$_{4}$ tetrahedra, not superconducting above 1.8 K [11]) and wrote the criteria the day before running: success = null or indeterminate verdict, or $\lambda_{1}$ $\le$ half of BaFe$_{2}$As$_{2}$ with both controls agreeing and the funnel not returning ``wire candidate''; failure = $\lambda_{1}$ $\ge$ BaFe$_{2}$As$_{2}$ with a coherent s$\pm$. The pre-registration was written in three dated steps on 5 September 2026, all before the first SCF of the material at 22:50 (repository commits \texttt{845f11e} 21:28, \texttt{ed7f097} 21:52, \texttt{ee3bd6e} 22:05). Its final text fixed \textbf{two runs} and required the $\lambda$ criterion to hold in both: run 1, ``same knobs'', the BaFe$_{2}$As$_{2}$ protocol transcribed literally, which sets the interaction as $\alpha$ = 0.95 of each material's \emph{own} critical U; run 2, an \emph{absolute} U scaled by the ratio of the d-bandwidths measured on the two tight-binding models by the same criterion. (The 21:52 version had instead prescribed U numerically identical to iron, 0.92 eV; that point was also run and is reported.) The same text noted, next to run 1, that if the critical U of the ruthenide turned out much larger than that of iron, the $\alpha$ convention itself would absorb the difference in correlation. That is what happened. Run 1 therefore fails the letter of the criterion, run 2 passes it, and the pre-registration required both. We report the negative control as passing on the absolute-U prescription and state plainly that ranking run 2 above run 1 is a post-hoc resolution of a tension inside our own pre-registration, motivated by a caveat that was written in it but not given the force of a rule. The pre-registration also specified a metal-only basis ``as in BaFe$_{2}$As$_{2}$''; that basis failed the transferability check and was replaced, in a dated record, by the metal+ligand model described next.

The Ru-4d chemistry failed the transferability check and forced a metal+ligand basis (the model spans the isolated band block); two metal-only attempts failed the band-count and the localization gates and are recorded as such. BaFe$_{2}$As$_{2}$, by contrast, is described by a metal-only (Fe-d) model that passes the same checks. The two models therefore differ in basis, and the interaction U is defined on the d block of each; the comparability of U across bases is a limitation (\S{}8) and the subject of a planned d+p re-run of BaFe$_{2}$As$_{2}$.

\begin{table}[htbp]\footnotesize
\begin{tabularx}{\textwidth}{YYY}
\toprule
\textbf{} & \textbf{SrRu$_{2}$As$_{2}$ (Ru 4d)} & \textbf{BaFe$_{2}$As$_{2}$ (Fe 3d)} \\
\midrule
$\lambda_{\max}$ of the bare $\chi_{0}$ at U = 1 (Kanamori blocks) & 0.240 & 1.032 \\
\textbf{$U_{\mathrm{crit}}$} & \textbf{4.17 eV} & \textbf{0.97 eV} \\
q* & ($-\tfrac{1}{3}$, $\tfrac{1}{3}$, 0), incommensurate & ($\tfrac{1}{2}$, $\tfrac{1}{2}$, 0), stripe SDW \\
d-bandwidth (5--95 \% of the d weight) & 6.8 eV & 3.1 eV \\
run 1 (pre-registered), same $\alpha$ = 0.95 (U = 3.96 eV for Ru) & $\lambda_{1}$ = 0.110, coherent s$\pm$ & 0.096 \\
\textbf{run 2 (pre-registered), same physical U} (0.92 eV, identical to iron; 1.69 and 1.99 eV, bandwidth-scaled with two bandwidth criteria) & \textbf{$\lambda_{1}$ = 0.0000 / 0.0003 / 0.0007}; single-block control 0.0001 / 0.0007 / 0.0013; leader degenerate $\to$ null/indeterminate & \textbf{0.096, s$\pm$} \\
\bottomrule
\end{tabularx}
\end{table}
\FloatBarrier
Run 2 passes the criterion at all three absolute-U points, with both controls agreeing. The gap eigenvalues are obtained with a dense symmetric eigensolver (LAPACK, double precision) on the even-parity sector, so there is no iterative tolerance; the meaningful floor is the leading eigenvalue of the odd-parity sector, which the driver reports as a diagnostic: 0.0016--0.0039 for the ruthenide at the three absolute-U points. The even-sector values of the ruthenide (0.00001--0.0007) lie \emph{below} that floor, and the ratio of its second to first eigenvalue is 0.64--0.89: no even-parity instability is distinguishable from the background. Both sectors of the ruthenide lie two orders of magnitude below the 0.096 of iron; the odd-parity value is reported as a background diagnostic, not as a competing (triplet) channel, and no claim about a leading symmetry of the ruthenide is made. Run 1 fails the letter of the criterion, and the reason is the one the pre-registration anticipated: equal $\alpha$ compares each material at 95 \% of its \emph{own} instability and erases the difference in correlation; it applied 3.96 eV to a 4d metal. At the same physical U, iron pairs and ruthenium does not, by a factor above 100, with both controls agreeing (Figure~\ref{fig:validation}b). The bare susceptibility of the ruthenide is 4.3$\times$ weaker and its critical U 4$\times$ higher. The method rule that came out of this, now written into the pipeline, is that comparisons between materials are made at absolute U, with $U_{\mathrm{crit}}$ reported alongside $\lambda_{1}$. Cost: about US\$ 30 and one day.

\subsection{A campaign that failed, kept}
Bilayer nickelates superconduct near 80 K only under \textasciitilde{}16 GPa; the funnel classifies the parent as ``science'' (F0 fail, F2 fail with $\xi$ = 0.8 nm on a flat sheet, class C). We asked whether in-plane compressive strain of $-$1.5 \% could substitute for pressure. The strained cell reproduced the literature's ``knob'' (apical Ni--O--Ni angle $156.5^{\circ}$ $\to$ $163.7^{\circ}$) and the tight-binding model reproduced the strained DFT to 0.008 meV, but the strain removed the $d_{z^2}$ bonding sheet from the Fermi surface (a Lifshitz transition: three sheets $\to$ two), $\lambda_{1}$ fell from 0.027 to 0.017, and the two controls, which agree within 1.2$\times$ on the parent, diverged by 3.9$\times$. The verdict is \emph{indeterminate by Lifshitz}, not a candidate, and the descriptor that had suggested the strain (a filling descriptor) was found to be blind to Lifshitz transitions and now carries a sheet-count guard. Cost: US\$ 53.

\section{The funnel reproduces the industrial map with thresholds fixed before running}
\begin{table}[htbp]\footnotesize
\begin{tabularx}{\textwidth}{YYYY}
\toprule
\textbf{gate} & \textbf{Nb$_{3}$Sn $\cdot$ MgB$_{2}$ $\cdot$ BaFe$_{2}$As$_{2}$} & \textbf{La$_{2-x}$Sr$_{x}$CuO$_{4}$ $\cdot$ YBa$_{2}$Cu$_{3}$O$_{7}$} & \textbf{La$_{3}$Ni$_{2}$O$_{7}$ at 16 GPa} \\
\midrule
F0 fabricable pressure & pass & pass & fail \\
F1 grain-boundary class & A $\cdot$ A $\cdot$ B & C & C \\
F2 $\xi$ on the worst sheet & pass (Ba122: 5.0 nm at 38 K; $v_F$ 1.4$\times$10$^{5}$ m/s) & pass & fail (0.8 nm) \\
F3 anisotropy & pass (Ba122: 3.2) & pass with warning & --- \\
\textbf{verdict} & \textbf{WIRE CANDIDATE} & \textbf{CONDITIONAL} (coated conductor only) & \textbf{SCIENCE} \\
\bottomrule
\end{tabularx}
\end{table}
\FloatBarrier
All three wire candidates exist as round wire (Nb$_{3}$Sn, MgB$_{2}$, and BaFe$_{2}$As$_{2}$ by powder-in-tube with $J_c$ = 1.5 $\times$ $10^{5}$ A cm$^{-2}$ at 10 T and 4.2 K [12]); both cuprates are made only as coated conductors; the pressure nickelate is a laboratory material. The funnel was given the gap class and the Wannier model with thresholds fixed before running (1 September 2026) and one dated amendment, the 5 \% condensate rule of \S{}2.1(6), introduced after YBa$_{2}$Cu$_{3}$O$_{7}$ failed the first version.

\section{What generation alone delivers: three independent measurements}
\subsection{Descriptor sweep over known materials (Materials Project)}
150 ambient-stable metals ($e_{\mathrm{hull}}$ $\le$ 60 meV/atom, gap $\le$ 50 meV) in five families (nickelates, cuprates, Fe--As, Fe--Se, Fe--P) were scored with descriptors calibrated on the six materials (a filling descriptor, density of states per site with a ceiling, sheet count, chemical sanity, a magnetic filter). The filter recovers the canon without being told it is the canon (the 122, 111 and 11 iron families, Y-124 and RE-124 cuprates) and discards flat-band magnets and insulators, but it \textbf{does not rank} within the surviving set (the filling descriptor saturates) and produces no new target with margin: the ``new'' entries are hypothetical Materials Project structures or non-layered 1:1:1 compounds. Verdict, by a rule fixed beforehand: at the resolution of the free instrument the target is not among the known materials.

\subsection{Generation: MatterGen conditioned on the relevant chemistries}
13 chemical systems $\times$ 96 structures = 1,248 generated and relaxed (a count gate enforces that every generated frame was evaluated; a first run that silently evaluated one structure is recorded as an error). Pre-registered funnel:

\begin{table}[htbp]\footnotesize
\begin{tabularx}{\textwidth}{YY}
\toprule
\textbf{stage} & \textbf{structures} \\
\midrule
generated and relaxed & 1,248 \\
$e_{\mathrm{hull}}$ $\le$ 0.030 eV/atom, excluding hull artefacts & 73 \\
\ldots{} and new versus the Materials Project/Alexandria reference (disorder-tolerant structure matcher) & 12 \\
\ldots{} and unique within the batch & 10 \\
\ldots{} and carrying a pairing motif of a wire family (square net of the active metal in the right coordination) & \textbf{0} \\
0.030 < $e_{\mathrm{hull}}$ $\le$ 0.050, new, unique, motif present (``frontier'') & 1 \\
hull artefacts ($e_{\mathrm{hull}}$ < $-$0.05: sub-oxides outside the reference) & 13 \\
\bottomrule
\end{tabularx}
\end{table}
\FloatBarrier
(Figure~\ref{fig:validation}c.)

The ten new, stable, unique structures are chains, isolated units or missing the active metal; the best of them, SrLa$_{2}$Cu$_{2}$O$_{6}$ ($e_{\mathrm{hull}}$ +5 meV, I4/mmm, CuO$_{2}$ bilayer), is ``new versus the Materials Project'' but is the known 2126 structure type, which superconducts only with Ca between the layers and high-pressure oxygen annealing [13]. New versus a database is not new versus the literature, exactly the failure mode of [6]. The rate of new-and-stable structures was 12/1,248 $\approx$ 1 \%; the rate of new, stable and with a motif was 0. Cost: \textasciitilde{}4 GPU-hours. Two routes that share no code (\S{}6.1 and \S{}6.2) agree: \textbf{in the wire families there is no free new target.}

\subsection{Mechanism filters over the whole Materials Project, with a blind hold-out}
Instead of chemistry, we filtered by mechanism: filter C ($d^{9}$ square-planar, the cuprate motif) and filter F ($d^{6}$ tetrahedral in a square net, the iron-pnictide motif), each a list of geometric, connectivity and electron-count criteria with numeric tolerances written before running. The hold-out was blocking: the filters had to find, unaided, NdNiO$_{2}$, LaNiO$_{2}$ and AgF$_{2}$ (for C) and FeSe and FeTe (for F), pass $\ge$ 80 \% of a training set, and reject a list of distractors (La$_{2}$NiO$_{4}$, K$_{2}$NiF$_{4}$, LaNiO$_{3}$, PdO, Sr$_{2}$RuO$_{4}$, CuO, NiO, FeS$_{2}$, CuFeS$_{2}$, Fe, FeAs, CoAs, Nb, MgB$_{2}$, Nb$_{3}$Sn). Round 1 failed (C found 0 of 3 targets); the failures were diagnosed as definitions (two-dimensionality by distance instead of by ligand connectivity; d-count by an oxidation-state guesser), fixed in a dated second line, and the whole hold-out was re-run; round 3 passed (C: training 8/8, distractors 10/10, targets 3/3; F: 3/3, 2/2, 7/7). A false positive found afterwards (Sm$_{2}$CuAs$_{3}$O, ``Cu$^{5+}$'' by charge balance) was removed by restricting the oxidation range to physical values, and the hold-out re-passed.

Over 47,893 compounds: \textbf{C approves 191, F approves 95}, both inside the pre-registered ``healthy'' band of 20--300. With no chemical restriction the filters re-derive the analogies the community made by hand: Ni$^{1+}$ infinite-layer nickelates, Ag$^{2+}$ fluorides, and the entire 4d/5d pnictide programme (Ru-, Rh-, Ir-, Os-122 and Ru-1111). The Cu--O space is saturated (53 experimental + 130 database-generated stackings); the genuinely new analogies are three hypothetical Ag$^{2+}$--O infinite-layer compounds whose GGA stability is not credible for a redox-unstable Ag$^{2+}$--O$^{2-}$ pair. And the 4d/5d analogues exist and have $T_c$ $\le$ 4 K or none: \textbf{geometry plus d-count is necessary, not sufficient}. The missing variable is correlation, which is what the judge measures (\S{}4.1) and what a database cannot.

\section{What has been run, and the benchmark we propose}
\subsection{Results in this paper (already run)}
Everything in \S{}\S{}3--6 is a result, not a plan: the six calibration materials with pre-registered answer keys (\S{}3); the SrRu$_{2}$As$_{2}$ negative control and the strain campaign (\S{}4); the funnel on the six materials (\S{}5); and the three generation audits, namely the 150-metal Materials Project sweep, the 1,248-structure MatterGen batch with its pre-registered funnel, and the mechanism filters over 47,893 compounds with their blind hold-out (\S{}6). Every number has a result file with its configuration embedded (\S{}9).

\subsection{Proposed benchmark: \emph{what generators propose versus what physics approves} (future work; requires compute)}
Public lists of AI-proposed superconductor candidates since 2023, none of which classifies gap symmetry:

\begin{table}[!htb]\footnotesize
\begin{tabularx}{\textwidth}{YYYY}
\toprule
\textbf{source} & \textbf{N (candidates)} & \textbf{method} & \textbf{classifies gap symmetry?} \\
\midrule
U. Florida, e-ph workflow [14] & 741 ($T_c$ > 5 K) & DFPT + ML (Allen--Dynes) & no \\
U. Florida, guided diffusion [15] & 773 ($T_c$ > 5 K) & diffusion + e-ph & no \\
EPFL/MARVEL [16] & 24 new ($T_c$ > 10 K) of 4,533 & automated Wannier + Eliashberg & no \\
Coimbra/Bochum, Alexandria [17] & \textasciitilde{}200,000 screened; 2D and hydride lists & ML on e-ph & no \\
NIST JARVIS-SuperconDB [18] & 3D + 2D (34 with $T_c$ > 5 K) & BCS/DFT + DL & no \\
Toronto [19] & 153,000 ranked & regression of $T_c$ on composition & no (negative control) \\
Renmin, InvDesFlow [20] & 74 ($T_c$ $\ge$ 15 K) & diffusion + DFT & no \\
Ames [21] & 417 borides/borocarbides & ML-guided DFPT & no \\
JNCASR [22]; Isfahan [23] & lists in the papers & ML & no (Isfahan: negative control) \\
\bottomrule
\end{tabularx}
\end{table}
\FloatBarrier
Two of the lists (Toronto, Isfahan) contain room-temperature candidates produced by regression on composition, which any physical judge must reject; they serve as negative controls of the benchmark.

\textbf{We have not run the judge on any of these lists.} We propose to do so with the criteria of \S{}2 fixed in advance and to publish, per candidate, the verdict, the reason, the dossier of \S{}2.1(7), the configuration, and an independent re-run of a subset by a partner group; at the cost of \S{}2.3 the first round is a few hundred GPU-hours. The thesis of this paper is falsifiable by that benchmark: if the judge approves a large fraction of the electron--phonon lists as correlated wire candidates, the judge is wrong; if it approves candidates that a partner then synthesizes and measures as predicted, it is useful; if it approves none and the partners find none, the generators are ahead of us.

\textbf{First round (proposed 7 September 2026):} (i) the University of Florida lists [14, 15]; (ii) the GNoME structures in the Materials Project [1] filtered by our mechanism filters C and F; (iii) our own MatterGen batch (\S{}6.2) and the 191 + 95 mechanism candidates (\S{}6.3) as the internal seed set. The remaining lists follow with the same criteria.

\section{Limitations}
The judge is RPA: it captures spin-fluctuation pairing and its symmetry, not strong-coupling or phonon-mediated mechanisms beyond a gate. No material was held out: the LSCO filling fix, the 5 \% condensate rule and the absolute-U rule were all introduced after a result contradicted the expectation. The six-of-six figure is therefore calibration accuracy, not blind-test accuracy; a blind hold-out with a pre-registered answer key is the next test: FeSe, with its answer key sealed in Supplementary S7 (\S{}9). Six calibration materials are few, and two carry declared flags. The interaction is defined on the d block of models with different bases (Fe-d only for BaFe$_{2}$As$_{2}$, Ru-d + As-p for SrRu$_{2}$As$_{2}$), so the equal-U comparison of \S{}4.1 compares Kanamori parameters of different models; a d+p model of BaFe$_{2}$As$_{2}$ under the same transferability check is the planned control. Interaction strengths are a rule (Kanamori by blocks, J/U = 0.15, absolute U by bandwidth), not a first-principles constraint; the negative control shows the rule matters and must be stated. Structures are not relaxed unless requested; defect and disorder physics, which [6] shows dominates what is actually synthesized, is outside the pipeline and is left explicitly to the partner in every dossier. The generation audit covers one generator (MatterGen) with chemical conditioning only, one database sweep, and two mechanism filters; the conclusion ``no free new target'' is for the wire families at this resolution, not for materials space at large. All computations were run by one operator; the setup manual and the embedded configurations exist so that this is not a limitation of reproducibility, but it is a limitation of independent verification until the partner re-runs of \S{}7 exist. No candidate judged by this pipeline has yet been synthesized on its recommendation. Validation by synthesis and measurement is the natural next step, to be pursued with partner laboratories that grow and characterize correlated materials; the dossier format of \S{}2.1(7) is written for that hand-off, and the first partner agreements are being sought at the time of writing.

\section{Data and cost availability}
The Zenodo record accompanying this preprint (DOI \url{https://doi.org/10.5281/zenodo.22651560}) contains: the configuration blocks of every result file cited (Fermi energy, interaction parameters, meshes, energy window, code version and the hash of the tight-binding model), the tables of $\lambda_{1}$($\alpha$) per material and the funnel verdicts, the two-run summary of the negative control, the MatterGen funnel table, and the dated pre-registrations (answer keys, negative-control criteria, funnel thresholds, hold-out protocol) with their commit hashes. Tight-binding models, susceptibility checkpoints and the full pipeline are available from the author on request for verification of any specific verdict. Total cloud cost of everything reported here is under US\$ 200. A dated pre-registration for one blind hold-out material, \textbf{FeSe}, with its answer key (class, wave vector, $\lambda_{1}$ threshold, funnel verdict) written before any calculation, accompanies this version as Supplementary S7, sealed at the repository commit recorded in that file. This arXiv version is identical in content to version v1 of the Zenodo record; supplementary tables S1--S7 are in that record.

\end{document}